\documentclass[10pt,a4paper]{article}
\title{boundary article}
\author{Adel Messaoudi, Régis Cottereau, Christophe Gomez}
\date{}

\usepackage[french,english]{babel}
\usepackage{amsthm,amsmath,amssymb,amsfonts}
\usepackage[utf8]{inputenc}
\usepackage{graphicx}
\usepackage{tikz,tkz-tab}
\usepackage{listings}
\usepackage{enumitem}
\usepackage{hyperref}
\usepackage{pdfpages}
\usepackage{bm}

\newcommand{ \B }{{ \bf b}}
\newcommand{ \BB }{{ \bf B}}
\newcommand{ \C}{ {\bf c}}
\newcommand{ \DD }{{ \bf D}}
\newcommand{ \ctilde }{\widetilde{c}}

\newcommand{\e}{ \mathrm{e} }

\newcommand{\ed}[1]{ a_{#1} }
\newcommand{\F}{ \mathcal{F} }
\newcommand{\I}{ \mathrm{i} }
\newcommand{ \K }{ {\bf k}}
\newcommand{ \Kperp }{ {\bf k}_\perp}
\newcommand{ \p }{ {\bf p}}
\newcommand{ \pperp }{ {\bf p}_\perp}
\newcommand{ \Pdiff }{ \mathcal{P}}
\newcommand{ \q }{ {\bf q}}

\newcommand{ \R }{ {\bf r}}

\newcommand{ \U }{ {\bf u}}
\newcommand{ \V }{ \widetilde{V}}
\newcommand{\W}{ \mathcal{W} }

\newcommand{ \x }{ {\bf x}}
\newcommand{ \xperp }{ {\bf x}_\perp}
\newcommand{ \xoeps }{ {\bf x}_{0,\epsilon}}
\newcommand{ \xn }{ x_n}
\newcommand{ \y }{ {\bf y}}
\newcommand{ \yperp }{ {\bf y}_\perp}

\usepackage[backend=bibtex]{biblatex}
\providecommand{\keywords}[1]
{
  \small	
  \textbf{Key words.} #1
}

\begin{document}

\title{Boundary effects in Radiative Transfer of acoustic waves in a randomly fluctuating half-space
}
\date{}
\author{Adel Messaoudi \footnotemark[1] $^,$\footnotemark[2]
\and Regis Cottereau \thanks{Aix-Marseille Univ, CNRS, Centrale Marseille, LMA, France.}
\and Christophe Gomez \thanks{Aix Marseille Univ, CNRS, I2M, France.}}

\maketitle
\numberwithin{equation}{section}

%\begin{tcbverbatimwrite}{tmp_\jobname_abstract.tex}
\begin{abstract}
This paper concerns the derivation of radiative transfer equations for acoustic waves propagating in a randomly fluctuating half-space in the weak-scattering regime, and the study of boundary effects through an asymptotic analysis of the Wigner transform of the wave solution. These radiative transfer equations allow to model the transport of wave energy density, taking into account the scattering by random heterogeneities. The approach builds on the method of images, where the half-space problem is extended to a full-space, with two symmetric sources and an even map of mechanical properties. Two contributions to the total energy density are then identified: one similar to the energy density propagation in a full-space, for which the resulting lack of statistical stationarity of the medium properties has no leading-order effect; and one supported within one wavelength of the boundary, which describes interference effects between the waves produced by the two symmetric sources. In the case of a homogeneous Neumann boundary conditions, this boundary effect yields a doubling of the intensity, and in the case of homogeneous Dirichlet boundary conditions, a canceling of that intensity.
\end{abstract}

\noindent \keywords{Radiative transfer, wave in random media, Wigner transform, boundary effects.}

%\begin{keyword}
%Radiative transfer, wave in random media, Wigner transform, boundary effects.
%\end{keyword}

%\begin{MSCcodes}
%74E35, 74J20, 74Q10, 82D30, 85A25
%\end{MSCcodes}
%\end{tcbverbatimwrite}
%\input{tmp_\jobname_abstract.tex}

%—————————————————————————————————————————
%INTRO
%—————————————————————————————————————————

\section{Introduction}
\label{sec:intro}

Radiative transfer theory was introduced over a century ago to describe the propagation of light in complex media. Today, it is used in many other fields such as in geophysics \cite{Maeda2008,Margerin2019,Margerin1998}, neutronics \cite{Larsen1975,Larsen1976,Larsen1974,Spanier2008}, optics \cite{Klose2010,Klose1999}, for weather forecasting \cite{Villefranque2019}, or for the illumination of animation movies scenes \cite{Bitterli2018,Keller2015}. Radiative transfer equations can be derived from the wave equation in the high-frequency regime \cite{Bal2005,Bal2010,Bal2002,Cottereau2014,butz2015,Ryzhik1996} through a multi-scale asymptotic analysis on the Wigner transform of the wave field. 

In this paper, we more particularly concentrate on (scalar) acoustics in a half-space $\Omega=\mathbb{R}^2 \times \mathbb{R}^\ast_-$. In that case, the wave equation for the pressure field $p(t,\x)$ is given by:
\begin{equation}
\label{equation des ondes}
\partial_{tt} p(t,\x)-c^2(\x)\Delta p(t,\x) = 0, \qquad \left(t, \x \right) \in \mathbb{R}_+^*\times \Omega,
\end{equation} 
where $c(\x)$ is the sound speed in the medium, modeled as a random field and assumed to fluctuate at the scale $\ell_c$ (correlation length) with amplitude $\sigma$ (standard deviation). Neumann boundary conditions complete this wave equation (homogeneous Dirichlet boundary conditions will also be considered further down):
\begin{equation}
\label{Neumann condition}
\partial_{n}p(t,\xperp,\xn=0)=0, \qquad \left(t, \xperp \right) \in \mathbb{R}_+^*\times \mathbb{R}^2 ,
\end{equation}
where the spatial variable $\x$ has been split into $\x = (\xperp,\xn)$, where $\xn$ represents the coordinate along the vector normal to the interface $\mathbb{R}^2 \times \{0\}$, $\xperp$ its transverse coordinate and $\partial_n$ stands for the derivative with respect to the variable $\xn$. Finally the following initial conditions are considered:
\begin{equation}\label{eq:initial_conditions_wave}
p(t=0,\x)=A(  \x-\x_{0} )\qquad \mathrm{and} \qquad \partial_t p(t=0,\x)=B( \x-\x_{0} ), \qquad \x\in\Omega.
\end{equation} 
where the shape of the functions $A(\x)$ and $B(\x)$ defines the wavelength $\lambda$. 

In the so-called high-frequency regime, where the parameter $\epsilon = \lambda/L$ is small ($L$ is the propagation length), and the medium velocity $c(\x)$ fluctuates weakly ($\sigma^2\approx\epsilon$) at the scale of the wavelength ($\ell_c\approx\lambda$), the energy density can be shown to verify a Radiative Transfer Equation (RTE):
\begin{equation}
\label{ETR}
\partial_t W +  c_0\widehat{\K}\cdot\nabla_{\x} W =- \Sigma(\K) W  +  \int_{\mathbb{R}^3}\sigma(\K,\q)W(\q) \delta(c_0(|\q|-|\K|))d\q, 
\end{equation}
where $W(t,\x,\K)$ represents the energy density at time $t$ and position $\x$ in direction $\K$ (see Sect.~\ref{High-frequency limit and Wigner transform} for a precise definition of Wigner transform), $\widehat{\K}=\K/|\K|$ is the normalized wave direction, and $c_0$ is the average sound speed in the medium. The differential scattering cross-section $\sigma(\K,\q)$ represents the rate at which energy density along wave vector $\q$ is diffracted into energy density along wave vector $\K$, and the total scattering cross-section is
\begin{equation}
\Sigma(\K) := \int_{\mathbb{R}^3}\sigma(\q,\K)\delta(c_0(|\q|-|\K|))d\q.
\end{equation}  

This asymptotic result is usually obtained for problems supported on the entire space ($\Omega=\mathbb{R}^3$), and has been derived starting from various wave equations and using different approaches (see references above). On the other hand, the case of the half-space ($\Omega=\mathbb{R}^2 \times \mathbb{R}^\ast_-$) has been less studied. The essential conclusion is that the RTE is still valid for half-spaces, and that it must be completed by a boundary condition on $\partial\Omega$, reminiscent of geometrical optics:
\begin{equation}
\label{reflecting boundary conditions}
W(t,\left(\xperp,\xn = 0), (\Kperp,k_n)\right) = W(t,\left(\xperp,\xn = 0), (\Kperp,-k_n)\right).
\end{equation}
This result has been obtained through direct plane wave analysis in the high-frequency regime~\cite{Jin2006b} (see also an extension to elastic waves~\cite{Jin2006, Savin2008}) or through more technical results on the trace of semiclassical measures~\cite{Gerard1993, Ryzhik1997, Burq1997, Miller2000} (see also the more complicated extension to the elastic case, treated in~\cite{Akian2012}). In~\cite{Bal1999}, the authors extend these results on flat interfaces to the case of rough interfaces (in the regime of Born approximation), and conclude that roughness induces additional diffraction at the interface. Finally, note that the heterogeneous half-space can also be modeled as a finite (heterogeneous) layer over a homogeneous half-space, in the limit of infinite thickness~\cite{Borcea2021}. This allows to work on the (countable) basis of modes of the finite-size layer, and shows coupling between the surface and bulk waves.

\begin{figure}[tbhp]
\label{Figure superposition ondes}
\centering \includegraphics[trim = 3.5cm 9cm 2.5cm 7cm,scale=0.15]{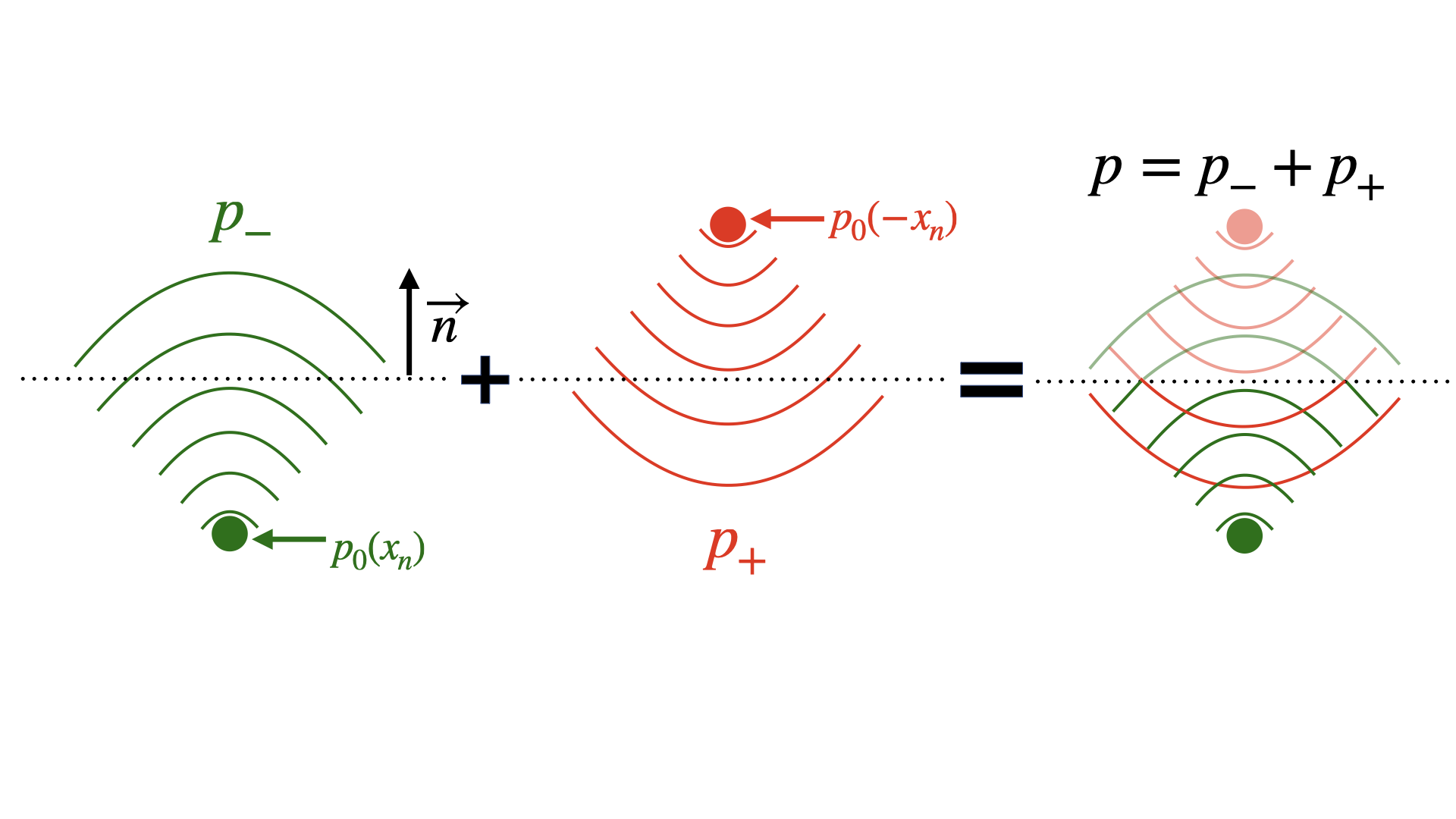}%modif ici
\caption{Sketch of the method of images, for the half-space bounded above by the dashed line: up-going wave $p_-$ propagating in a full-space from the real source (in green); down-going wave $p_+$ propagating from the fictitious source (in red); and real wave propagating from the real source in the half-space (rightmost figure).}
\label{fig:testfig}
\end{figure}

The first objective of this paper is to retrieve the result that high-frequency wave energy propagation in a half-space are controlled by a RTE, Eq.~\eqref{ETR}, with the boundary condition \eqref{reflecting boundary conditions}, using the method of images. This approach  allows the study of boundary effects in a convenient way. The idea of the method of images is to use the equivalence of the solution $p(t,\x)$ of \eqref{equation des ondes}, supported on the half-space $\Omega$ with the superposition of two solutions $p_-(t,\x)$ (up-going wave) and $p_+(t,\x)$ (down-going wave) of the same wave equation~\eqref{equation des ondes} extended to the full-space $\mathbb{R}^3$, the former with the original initial condition, and the latter with an initial condition chosen so as to enforce the boundary condition (see Fig.~\ref{Figure superposition ondes} for a sketch). The extension of the original wave equation~\eqref{equation des ondes} to the full-space is done such that the velocity field $c(\xperp,\xn)$ is an even function of $\xn$. This approach has the advantage that it works on the full-space, bypassing the need to deal specifically with the interface, as in previous papers. However, since the pressure field is decomposed in two parts, four components of the Wigner transform will have to be considered. Also, the extension of the random velocity field induces a loss of statistical homogeneity, but it will be shown that our approach provides no first-order influence on the RTE (Sect.~\ref{Radiative transfer equation}) and that the classical results for a full-space are retrieved. 

Another objective of this paper, not previously considered in the literature, is to consider more in detail what happens at the interface. More specifically, we concentrate on what happens at a distance of the order of the wavelength $\lambda$ (scaling as $\epsilon$) from the interface $\xn=0$. At this scale, interference phenomena occur which translate in a doubling of the total energy with respect to classical RTE with Neumann boundary conditions (and a cancellation in the case of Dirichlet boundary conditions). The contribution due to the interference effects vanishes far from the border of the propagation medium, except when the initial condition is close (within one wavelength) to the boundary. In that case, the up and down-going waves evolve within one wavelength of each other and then produce constructive interferences throughout the propagation domain.

The outline of this paper is as follows. In Sect.~\ref{Acoustic waves in a random half space}, the method of images is described in detail, as well as the asymptotic regime that is considered and the Wigner transform. Sect.~\ref{Asymptotic analysis with a source far from the border} describes the main results of this paper, deriving the RTE on the basis of the method of images, as well as exhibiting the constructive interference phenomenon at the boundary, for a source far from the boundary. Finally, Sect.~\ref{Asymptotic analysis with a source close to the border} addresses the case when the initial condition is supported close to the border.

%—————————————————————————————————————————
%Acoustic waves in a random half-space in the high-frequency regime
%—————————————————————————————————————————

\section{Acoustic waves in a random half-space in the high-frequency regime}
\label{Acoustic waves in a random half space}

In the following, we consider the scalar wave equation~\eqref{equation des ondes} over the half-space $\Omega=\mathbb{R}^2 \times \mathbb{R}^\ast_-$, equipped with the boundary condition~\eqref{Neumann condition} and the initial conditions~\eqref{eq:initial_conditions_wave}. It is assumed that the initial conditions $A$ and $B$ are smooth functions on $\mathbb{R}^3$, even and compactly supported w.r.t. the $x_n$-variable to be compatible with the boundary conditions. %For instance, with Neumann boundary conditions, Eq.~\eqref{Neumann condition}, we set \textcolor{red}{$\partial_{n}A(\xperp,\xn=0)=0$.}

%—————————————————————————————————————————
%The method of images
%—————————————————————————————————————————

\subsection{The method of images} 
\label{The method of images}

The basic principle of the method of images is to replace an acoustic problem posed on a half-space with a given initial condition (or source) and velocity field by an acoustic problem posed on the full-space with a symmetrized initial condition and velocity field so that the solution of these problems are the same on the original half-space. We therefore consider the following extension $\ctilde(\x)$ to the full-space of the original velocity field $c(\x)$ defined only over the (lower) half-space:
\begin{equation}
\label{sound speed extension}
\ctilde(\x):=c(\xperp,\xn)\mathbf{1}_{\mathbb{R}_-}(\xn) +  c(\xperp,-\xn)\mathbf{1}_{\mathbb{R}_+}(\xn), \quad \x\in\mathbb{R}^3.
\end{equation}  
Note that $\ctilde$ is an even function with respect to the $x_n$-variable.  We then consider a pair of solutions (in the full-space $\mathbb{R}^3$) denoted up-going wave $p_-(t,\x)$ and down-going wave $p_+(t,\x)$, to the same equation
\begin{equation}
\label{equation des onde full-space}
\partial^2_{tt} p_\pm(t,\x)-\ctilde^2\left(\x \right) \Delta p_\pm(t,\x) = 0 \qquad  \left(t, \x \right) \in \mathbb{R}_+^*\times \mathbb{R}^3,
\end{equation}   
but with different initial conditions:
\begin{equation}\label{eq:IC_p-+}
\begin{cases}
p_-(t=0,\x) = A( \xperp-\x_{0\perp},\xn-x_{0n} ), \\
p_+(t=0,\x) = A( \xperp-\x_{0\perp},\xn+x_{0n} ), 
\end{cases}
\end{equation}
and
\begin{equation}\label{eq:IC_dp-+}
\begin{cases}
\partial_tp_-(t=0,\x) = B ( \xperp-\x_{0\perp},\xn-x_{0n} ), \\
\partial_tp_+(t=0,\x) = B ( \xperp-\x_{0\perp},\xn+x_{0n} ).
\end{cases}
\end{equation}
Finally, we introduce $\widetilde{p}$: 
\begin{equation}
\label{superposition}
\widetilde{p}(t,\x) := p_-(t,\x) + p_+(t,\x), \quad (t,\x) \in  \mathbb{R}_+\times \mathbb{R}^3.
\end{equation}
The choices above induce the following symmetry of the up-going and down-going fields
\begin{equation}\label{sym_p-p+}
 p_+(t,\xperp,\xn) = p_-(t,\xperp,-\xn), \quad (t,\xperp,\xn) \in  \mathbb{R}_+\times \mathbb{R}^2\times\mathbb{R},
\end{equation}
which ensures that the derivative $\partial_n\widetilde{p}(t,\x)$ vanishes for $\xn=0$, and that the restriction of $\widetilde{p}(t,\x)$ to the lower half-space is equal to the solution of the original problem:
\begin{equation}\label{def:p_field}
\widetilde{p}(t,\x) = p(t,\x), \quad \left(t, \x \right) \in \mathbb{R}_+^*\times \Omega.
\end{equation}
The derivations above are appropriate for Neumann conditions on the interface $\xn=0$. If Dirichlet boundary conditions were considered, the sign of $p_+(t,\x)$ in \eqref{def:p_field} should be switched to a $-$ sign. This ensures again that the restriction of $\widetilde{p}(t,\x)$ to the lower half-space is equal to the solution of the original problem $p(t,\x)$.

%—————————————————————————————————————————
%High-frequency regime
%—————————————————————————————————————————

\subsection{High-frequency regime}

The high-frequency regime is obtained by considering the scaling $t \to t/\epsilon,$ $\x \to \x/\epsilon$ for a small parameter $\epsilon = \lambda/L\ll 1$. We thus set
\begin{equation}
p^\epsilon(t,\x) := p\left(\frac{t}{\epsilon},\frac{\x}{\epsilon} \right) \quad \left(t, \x \right) \in \mathbb{R}_+^*\times \Omega,
\end{equation}
and, in a similar fashion, the rescaled $p^\epsilon_-(t,\x)$, $p^\epsilon_+(t,\x)$, and $\widetilde{p}^\epsilon(t,\x)$ associated respectively to $p_-$, $p_+$, and $\tilde p$. Since the orders of derivation in time and space are the same for wave equations, the equilibrium equations~\eqref{equation des ondes} and~\eqref{equation des onde full-space}, for $p^\epsilon$, $p^\epsilon_-$ and $p^\epsilon_+$ are unchanged. The initial conditions are rescaled as
\begin{equation}\label{eq:cond_init_eps}
p^\epsilon(t=0,\x)=\frac{1}{\epsilon^{3/2}}A\left( \frac{ \x-\x_{0} }{\epsilon}  \right)\qquad\text{and}\qquad \epsilon \, \partial_t p^\epsilon(t=0,\x)=\frac{1}{\epsilon^{3/2}}B\left(  \frac{ \x-\x_{0} }{\epsilon}  \right),
\end{equation}
where the amplitudes are chosen so as to provide solutions of the RTE of order 1, and similarly for the initial conditions~\eqref{eq:IC_p-+} and~\eqref{eq:IC_dp-+}. The velocity field is assumed to fluctuate around a constant background value in the following manner (a slowly fluctuating background can also be considered, see for instance~\cite{Ryzhik1996}):
\begin{equation}
\label{sound speed}
c^2\left( \frac{\x}{\epsilon} \right):= c_0^2 - \sqrt{\epsilon} \, V \left(\frac{\x}{\epsilon}\right), \quad \x\in \Omega,
\end{equation}
where $c_0$ is the (constant) background velocity and $V(\x)$ accounts for the random fluctuations. The scaling $\sqrt{\epsilon}$ is the proper scaling that allows the random fluctuations to fully interact with the waves and to provide, after an asymptotic analysis, a RTE describing the energy propagation. 

Although it is possible to work directly with the second-order form of the wave equation to derive the RTE, we use the formalism described in~\cite{Bal2005}. We therefore introduce the vector field 
\begin{equation}
\label{equation des ondes ordre 1}
\U^\epsilon(t,\x):=\begin{pmatrix} \widetilde{p}^\epsilon(t,\x)   \\ \epsilon\,  \ctilde^{-2}\left(\frac{\x}{\epsilon}\right)   \partial_t \widetilde{p}^\epsilon(t,\x) \end{pmatrix}, \quad (t,\x)\in  \mathbb{R}_+\times\mathbb{R}^3.
\end{equation}
Through the relation \eqref{superposition}, $\U^\epsilon$ is actually composed of a sum of two vectors:
\begin{equation}
\label{superposition vecteurs}
\U^\epsilon(t,\x)=\U^\epsilon_-(t,\x) + \U^\epsilon_+(t,\x),
\end{equation}
and, for all $j\in\left\{-,+\right\},$ $\U^\epsilon_j$ satisfies the $2\times2$ system of equations
\begin{equation}
\label{equation hyperbolique ordre 1}
\epsilon\partial_t \U^\epsilon_j + \mathcal{A}_\epsilon\U^\epsilon_j = 0, \qquad \mathrm{ where }\qquad \mathcal{A}_\epsilon:=-\begin{pmatrix}0& \ctilde^2 \left( \frac{\x}{\epsilon} \right)\\ \epsilon^2\Delta & 0  \end{pmatrix}, %+ \sqrt{\epsilon}V\left( \frac{\x}{\epsilon} \right)K ,
\end{equation}
with appropriate initial conditions inherited from~\eqref{eq:IC_p-+} and~\eqref{eq:IC_dp-+}.

%—————————————————————————————————————————
%Structure of the inhomogeneities
%—————————————————————————————————————————

\subsection{Structure of the inhomogeneities}\label{Structure of the inhomogeneities}

The fluctuation field $V(\x)$ is modeled as the restriction on $\Omega$ of a statistically homogeneous mean-zero random field defined for all $\x\in\mathbb{R}^3$, with (normalized) power spectrum density $\widehat{R}(\p)$ given by
\begin{equation*}
%\label{power spectra half space}
(2\pi)^3c_0^4\widehat{R}(\p)\delta(\p+\q):=\big\langle\widehat{V}(\p)\widehat{V}(\q)\big\rangle,
\end{equation*}
such that 
\begin{equation}\label{eq:R_hat}
\widehat{R}(-\p)=\widehat{R}(\p).
\end{equation}
Here, $\langle\cdot\rangle$ denotes an ensemble average and $\widehat{V}(\K)$ refers to the Fourier transform of $V(\x)$, with the convention
\begin{equation*}
\widehat{V}(\K) := \int_{\mathbb{R}^3} \e^{ - \I \K \cdot \x} V(\x) d\x \qquad \mathrm{and} \qquad V(\x) := \frac{1}{(2\pi)^3}\int_{\mathbb{R}^3} \e^{  \I \K \cdot \x}\widehat{V}(\K) d\K.
\end{equation*}
Note that the property \eqref{eq:R_hat} is satisfied if we assume the inverse Fourier transform $R$ of $\widehat{R}$ to be of the form $R(\x)=r(|\x|)$, $R$ being therefore the correlation function of $V$. Let us also remark that due to the even extension of the velocity field properties w.r.t. the $x_n$-variable (Eq.~\eqref{sound speed extension}) the corresponding random fluctuations $\V(\x/\epsilon)$, defined as
\begin{equation*}
\V\Big(\frac{\x}{\epsilon}\Big):=V\Big(\frac{\xperp}{\epsilon},\frac{\xn}{\epsilon}\Big)\mathbf{1}_{\mathbb{R}_-}(\xn) +  V\Big(\frac{\xperp}{\epsilon},-\frac{\xn}{\epsilon}\Big)\mathbf{1}_{\mathbb{R}_+}(\xn),
\end{equation*}
in the high-frequency scaling introduced above, are not stationary anymore w.r.t. this variable. Nevertheless, the asymptotic analysis we provide in Sect. \ref{sec:multi_scale} is based on a separation of the slow and fast $\x$-components, for which $\V$ can then be recast as
\begin{equation*}
\V\left(\frac{\x}{\epsilon}\right)=\V(\x,\y)_{|\y = \frac{\x}{\epsilon}}=V(\y_\perp,y_n)\mathbf{1}_{\mathbb{R}_-}(\xn) +  V(\y_\perp,-y_n)\mathbf{1}_{\mathbb{R}_+}(\xn)\, {}_{|\y = \frac{\x}{\epsilon}},
\end{equation*}
and where the slow and fast variable are separated. The slow variable being only produced by the extension procedure. In this way, for any fixed slow component $\x$, the power spectrum density to $V(\x,\y)$ w.r.t. the fast component reads
\begin{equation}
\label{Power spectra}
\Big\langle \widehat{\V}(\x,\p)  \widehat{\V}(\x,\q)\Big \rangle= (2\pi)^3c_0^4\delta(\p+\q)\widehat{R}(\p),
\end{equation}
where $\widehat{\V}(\x,\p)$ stands for the Fourier transform with respect to the $\y$-variable. Note that the obtained power spectra does not depend on the slow component $\x$, so that the extension procedure will play no role in the limiting RTE derived in Sect. \ref{Asymptotic analysis with a source far from the border}.

%—————————————————————————————————————————
%Wigner transform
%—————————————————————————————————————————

\subsection{Wigner transform}
\label{High-frequency limit and Wigner transform}

%Radiative transfer models arise in the high-frequency limit of wave propagation, we thus set 
%\begin{equation}
%p^\epsilon_j(t,\x) = p_j\left(\frac{t}{\epsilon},\frac{\x}{\epsilon} \right),
%\end{equation}
%and the following equations are considered 
%\begin{equation}
%\label{equation des ondes HF}
%\partial_{tt} p^\epsilon_j(t,\x)-\widetilde{c}_\epsilon^2\left(\frac{\x}{\epsilon}\right) \Delta p^\epsilon_j(t,\x)  = 0, \quad (t,\x)\in \mathbb{R}_+^*\times \mathbb{R}^3,
%\end{equation}
%for $j\in \left\{-,+\right\}$. These equation are equipped  with initial conditions
%\begin{equation}
%p_-^\epsilon(t=0,\x)=\frac{1}{\epsilon^{3/2}}A\left( \frac{ \x-\x_{0}}{\epsilon}   \right)\quad \mathrm{and} \quad \partial_t p_-^\epsilon(t=0,\x)=\frac{1}{\epsilon^{3/2}}B\left( \frac{ \x-\x_{0}}{\epsilon}  \right),
%\end{equation} 
%the symmetry relation \eqref{sym} allows to deduce $p_+^\epsilon(t=0,\x)$ and $\partial_t p_+^\epsilon(t=0,\x).$ Let us now introduce $p_\epsilon$ defined by 
%\begin{equation}
%p^\epsilon(t,\x) = p_-^\epsilon(t,\x)+p_+^\epsilon(t,\x),
%\end{equation}
%this scalar wave field can be viewed as a rescaled version of $\widetilde{p}$ in the high-frequency regime.
The derivation of RTE from the wave equation in the high-frequency regime relies on a multi-scale asymptotic analysis of the Wigner transform of the wave field. The Wigner transform of two vector fields $\mathbf{v}$ and $\mathbf{w}$ is defined as
\begin{equation*}
W[\mathbf{v},\mathbf{w}](\x,\K):=\int_{\mathbb{R}^3}\e^{\I\K\cdot \y} \mathbf{v}\left(\x- \frac{\epsilon\y}{2}\right)\mathbf{w}^\ast\left(\x+ \frac{\epsilon \y}{2}\right)\frac{d\y}{(2\pi) ^3}.
\end{equation*}
Here $\mathbf{w}^\ast$ means transposition of the complex conjugate of vector $\mathbf{w}$. We may think of the Wigner transform as the inverse Fourier transform of the two point correlation function of $\mathbf{v}$ and $\mathbf{w}$.

Let us consider $\U^{\epsilon}$ defined by \eqref{equation des ondes ordre 1} with Wigner transform 
\begin{equation}
\label{Wigner u}
W_\epsilon(t,\x,\K) := W[\U^{\epsilon}\left( t, \cdot \right),\U^{\epsilon}\left( t, \cdot \right)](\x,\K), \quad \left(t,\x,\K\right)\in\mathbb{R}_+\times\mathbb{R}^3\times\mathbb{R}^3.  
\end{equation}
Since we are using the method of images, according to \eqref{superposition vecteurs}, four Wigner transforms actually have to be accounted for: 
\begin{equation}
\label{Wigner total}
W_\epsilon(t,\x,\K) = W_\epsilon^{--}(t,\x,\K) + W^{++}_\epsilon(t,\x,\K) +W^{-+}_\epsilon(t,\x,\K)+ W_\epsilon^{+-}(t,\x,\K),  
\end{equation}
where 
\begin{equation*}
%\label{Wigner terme a terme}
W_\epsilon^{ij}(t,\x,\K):= W[\U_i^{\epsilon}\left( t, \cdot \right),\U_j^{\epsilon}\left( t, \cdot \right)](\x,\K),\quad \forall i,j\in\{-,+\}.
\end{equation*}
However, using the symmetry relation \eqref{sym_p-p+} between the up-going and down-going waves, we observe:
\begin{equation}
\label{sym wigner_self}
W_\epsilon^{- -}(t,\x,\K)= W_\epsilon^{++}\left(t,(\xperp,-\xn),(\Kperp,-k_n)\right), \quad \left(t,\x,\K\right)\in\mathbb{R}_+\times\mathbb{R}^3\times\mathbb{R}^3,
\end{equation} 
and 
\begin{equation}
\label{sym wigner_cross}
W_\epsilon^{- +}(t,\x,\K)= W_\epsilon^{+-}\left(t,(\xperp,-\xn),(\Kperp,-k_n)\right), \quad \left(t,\x,\K\right)\in\mathbb{R}_+\times\mathbb{R}^3\times\mathbb{R}^3.
\end{equation} 
Hence it is only necessary to follow two Wigner transforms and we will focus for the remaining of the paper on $W_\epsilon^{- -}(t,\x,\K)$ and $W_\epsilon^{- +}(t,\x,\K)$, that we will denote respectively as self-Wigner transform and cross-Wigner transform.

%—————————————————————————————————————————
%Asymptotic analysis with an initial condition far from the border
%—————————————————————————————————————————

\section{Asymptotic analysis with an initial condition far from the border}
\label{Asymptotic analysis with a source far from the border}

In this section, models for energy density propagation carried by the up and down-going waves (actually only $W_\epsilon^{- -}$) are derived through an asymptotic analysis in the context of an initial condition \emph{far} from the border. Here, by \emph{far} we mean that most of the mass of the initial condition is located at a distance of order $1$ from the border compared to the wavelength which is of order $\epsilon$. The intensity enhancement happening at the border of the domain $\partial \Omega$ is provided by nontrivial limits for the cross-terms (only $W_\epsilon^{- +}$ will be considered), which can be explicitly described through the asymptotic behavior of $W_\epsilon^{- -}$ as we will see later. For an initial condition located close to the border, the intensity enhancement phenomena are different from what will be described below, and will be precisely treated in Sect. \ref{Asymptotic analysis with a source close to the border}. 

%—————————————————————————————————————————
%Radiative transfer equations for the self-Wigner transforms
%—————————————————————————————————————————

\subsection{Radiative transfer equations for the self-Wigner transforms} 
\label{RTE for the self-Wigner transforms}
The derivation of the radiative transfer model for $W_\epsilon^{- -}$ is based on a multi-scale asymptotic analysis and can be done rigorously in our context following \cite{butz2015}. This approach would provide a stronger convergence result, an almost sure convergence in some weak sense for $W_\epsilon^{- -}$ as $\epsilon\to 0$ to a deterministic limit, but its methodology is beyond the scope of this paper. In this paper we rather follow the approach proposed in \cite{Bal2005}, that we remind in Sect. \ref{sec:multi_scale} and \ref{Radiative transfer equation} for the sake of completeness. Let us remark that the later approach provides only a formal convergence of $\langle W_\epsilon^{- -}\rangle$, but it still brings the main characteristics of the asymptotic radiative transfer model. 

%—————————————————————————————————————————
%Equation for the Wigner transform
%—————————————————————————————————————————

\subsubsection{Equation for the Wigner transform}

From \eqref{equation hyperbolique ordre 1} the following relation holds true 
\begin{equation}
\label{eq Wigner}
\epsilon \partial_t W_\epsilon^{- -}(t,\x,\K)  + W\left[\mathcal{A}_\epsilon\U^\epsilon_-(t,\cdot),\U^\epsilon_-(t,\cdot) \right](\x,\K) \\
+ W\left[\U^\epsilon_-(t,\cdot),\mathcal{A}_\epsilon\U^\epsilon_- \left(t,\cdot\right)\right](\x,\K) =0,
\end{equation}
for $(t,\x,\K)\in  \mathbb{R}_+^\ast \times \mathbb{R}^3\times\mathbb{R}^3$, where according to \eqref{sound speed}, the operator $\mathcal{A}_\epsilon$ will be split as
\begin{equation*}
\mathcal{A}_\epsilon=-\begin{pmatrix}0 & c_0^2 \\  \F\left( \epsilon \mathbf{D} \right)& 0  \end{pmatrix} + \sqrt{\epsilon}\, \V\left(\x, \frac{\x}{\epsilon} \right)K \qquad\text{with}\qquad K =\begin{pmatrix}0& 1\\ 0& 0  \end{pmatrix}.
\end{equation*}
In $\mathcal{A}_\epsilon$ the Laplacian operator is replaced by the pseudo-differential operator $\F\left( \epsilon \mathbf{D} \right)$ defined by
\begin{equation*}
\F\left( \epsilon \mathbf{D} \right)\left[\U\right](\x)=\int_{\mathbb{R}^3} \e^{\I\p \cdot \x} \left(\I\epsilon\p \right) \cdot \left(\I\epsilon\p \right) \widehat{\U}(\p)\frac{d\p}{(2\pi)^3}.
\end{equation*}
Knowing the four following relations  
\begin{align*}
W\left[\F(\epsilon \mathbf{D})\U^\epsilon_-(t,\cdot),\U^\epsilon_-(t,\cdot) \right](\x,\K) & = \F\left( \I\K + \frac{\epsilon \mathbf{D}}{2} \right)\left[W_\epsilon^{--}\right](t,\x,\K),\\
W\left[\U^\epsilon_-(t,\cdot),\F(\epsilon \mathbf{D})\U^\epsilon_- (t,\cdot)\right](\x,\K) & = \F\left( \I\K - \frac{\epsilon \mathbf{D}}{2} \right)\left[W_\epsilon^{--}\right](t,\x,\K),\\
W\left[\V\left( \x,\frac{\x}{\epsilon} \right)\U^\epsilon_-(t,\cdot),\U^\epsilon_- (t,\cdot)\right](\x,\K) & = \int_{\mathbb{R}^3}\e^{\I \x \cdot\p/\epsilon}\widehat{\V}(\x,\p)W^{- -}_\epsilon\left(t,\x,\K-\frac{\p}{2}\right)\frac{d\p}{(2\pi)^3} \\
& + O\left( \epsilon \right),\\
W\left[\U^\epsilon_-(t,\cdot),\V\left( \x,\frac{\x}{\epsilon} \right)\U^\epsilon_- (t,\cdot)\right](\x,\K)& = \int_{\mathbb{R}^3}\e^{\I \x \cdot\p/\epsilon}\widehat{\V}(\x,\p)W^{- -}_\epsilon\left(t,\x,\K+\frac{\p}{2}\right)\frac{d\p}{(2\pi)^3} \\
& + O\left( \epsilon \right) ,
\end{align*}
Eq. \eqref{eq Wigner} can be recast as
\begin{multline}
\label{Wigner equation}
\epsilon \partial_t W_\epsilon^{- -}(t,\x,\K) + \Pdiff\left(\I \K+ \frac{\epsilon\mathbf{D}}{2}    \right) W_\epsilon^{--}(t,\x,\K)+ 
 W_\epsilon^{--}(t,\x,\K) \Pdiff^\ast \left(\I \K- \frac{\epsilon\mathbf{D}}{2}    \right) \\
+ \sqrt{\epsilon}\left(K\, \mathcal{K}^-_\epsilon  \left[W_\epsilon^{- -}\right](t,\x,\K) +  \mathcal{K}_\epsilon^+\left[W_\epsilon^{- -}\right](t,\x,\K) \, K^\ast \right) + O( \epsilon^{3/2}) =0,
\end{multline}
where
\begin{equation}
\label{def:calP}
\Pdiff\left(\I \K+ \frac{\epsilon\mathbf{D}}{2}    \right) :=-\begin{pmatrix}  0 & c_0^2\\ \F\left(\I\K+ \frac{\epsilon\mathbf{D}}{2}    \right)&0  \end{pmatrix},
\end{equation}
and
\begin{equation*}
\mathcal{K}_\epsilon \left[W\right](t,\x,\K)  := \int_{\mathbb{R}^3}\e^{\I \x \cdot\p/\epsilon}\widehat{\V}(\x,\p)W\left(t,\x,\K -\frac{\p}{2}\right)\frac{d\p}{(2\pi)^3}. 
\end{equation*}
Note that \eqref{Wigner equation} is naturally equipped with an initial condition at $t = 0$ depending directly on the ones of $\U^\epsilon_-$. At the limit $\epsilon \to 0$ this condition is given by
\begin{equation}
\label{eq:cond_init}
\lim_{\epsilon\to 0} W_\epsilon(t=0,\x,\K) = \widehat{S}(\K) \widehat{S}^\ast(\K) \delta(\x-\x_0),\qquad\text{with}\qquad  \widehat{S}(\K) := \begin{pmatrix} \widehat{A}(\K) \\  c_0^{-2}\widehat{B}(\K) \end{pmatrix}.  
\end{equation} 

%—————————————————————————————————————————
%Multiple scale expansion
%—————————————————————————————————————————

\subsubsection{Multiple scale expansion}
\label{sec:multi_scale} 

Due to the presence of the rapidly oscillating phases $\e^{\I \x \cdot\p/\epsilon}$ in $\mathcal{K}_\epsilon$, the fast variable $\y=\x /\epsilon$ is introduced and $W_\epsilon^{- -}$ is rewritten as
\begin{equation*}
W_\epsilon^{- -}(t,\x,\K)  = W_\epsilon^{- -}(t,\x,\y,\K)_{|\y=\x /\epsilon}=W_\epsilon^{- -}\left(t,\x,\frac{\x}{\epsilon},\K\right).
\end{equation*}   
%\begin{equation}
%W_\epsilon^{- -}(t,\x,\K) =W_\epsilon^{- -}\left(t,\x,\frac{\x}{\epsilon},\K\right) = W_\epsilon^{- -}(t,\x,\y,\K)
%\end{equation}
to account for this new variable. Having two spatial variables for $W_\epsilon^{- -}$, the differential operator $\mathbf{D}$ is now given by
\[
\mathbf{D} = \mathbf{D}_\x + \frac{1}{\epsilon}\mathbf{D}_\y, 
\]
and \eqref{Wigner equation} can be rewritten as
\begin{multline}
\label{Wigner equation rescale}
\epsilon \partial_t W_\epsilon^{- -} (t,\x,\y,\K)  + \Pdiff\left(\I\K+ \frac{\mathbf{D}_\y}{2} + \frac{\epsilon\mathbf{D}_\x}{2}    \right)W_\epsilon^{- -} (t,\x,\y,\K) \\
 + W_\epsilon^{- -} (t,\x,\y,\K) \Pdiff^\ast \left(\I\K - \frac{\mathbf{D}_\y}{2} - \frac{\epsilon\mathbf{D}_\x}{2}    \right)\\
+ \sqrt{\epsilon}\left( K \, \mathcal{K}  [W_\epsilon^{- -}](t,\x,\y,\K) +  \mathcal{K}^\ast [W_\epsilon^{- -}](t,\x,\y,\K) \, K^\ast \right) +O(\epsilon^{3/2})=0,
\end{multline}
with 
\begin{equation*}
\mathcal{K} [\W](t,\x,\y,\K):=\int_{\mathbb{R}^3}e^{i\y\cdot\p}\widehat{\V}(\x,\p)\W\left(t,\x,\y,\K - \frac{\p}{2}\right)\frac{d\p}{(2\pi)^3}.
\end{equation*}

To derive the radiative transfer equation from \eqref{Wigner equation rescale}, we consider the following expansion for $W_\epsilon^{- - }$ in powers of $\epsilon$
\begin{equation}
\label{asymptotic expansion 1}
W_\epsilon^{- - }(t,\x,\y,\K) = W_0(t,\x,\K) + \sqrt{\epsilon}\, W_1(t,\x,\y,\K) + \epsilon \, W_2(t,\x,\y,\K),
\end{equation} 
so that the asymptotic behavior of $W_\epsilon^{- - }$ is characterized by $W_0$. We also consider the first order expansion
\begin{equation}
\label{asymptotic expansion0}
\F\left(\I\K+ \frac{\mathbf{D}_\y}{2} + \frac{\epsilon\mathbf{D}_\x}{2}    \right) = \F\left(\I\K+ \frac{\mathbf{D}_\y}{2}  \right) + \frac{\epsilon}{2}\, \F'\left(\I\K+ \frac{\mathbf{D}_\y}{2}  \right) \cdot\nabla_\x + \mathrm{O}(\epsilon^2),
\end{equation}
where the symbol of $\F'$ is given by
\begin{equation*}%\label{def:F'}
\F'(\I \K) := -2\K = 2\I q_0(\I \K)\nabla_\K q_0(\I\K),
\end{equation*}
and the one of $q_0$ is defined as  
\begin{equation*}%\label{def:q0}
q_0(\I \K):=\sqrt{-\mathcal{F}(\I \K)} = |\K|.
\end{equation*}
This later symbol is introduced for notational conveniences in what follows. Remembering \eqref{def:calP}, $\Pdiff$ also admits an expansion of the form 
\begin{equation}
\label{asymptotic expansion}
\Pdiff=\Pdiff_0 + \epsilon\, \Pdiff_1 + \mathrm{O}(\epsilon^2),
\end{equation} 
where each term is defined according to \eqref{asymptotic expansion0}. Injecting \eqref{asymptotic expansion 1} and \eqref{asymptotic expansion} into \eqref{Wigner equation rescale} yields a sequence of three equations by equating the coefficients associated to each power of~$\epsilon$.

% ----------------------------------------------------------------------
  
\paragraph{Leading order term $W_0$ and dispersion relation}

The leading order terms yield the relation
\begin{equation}
\label{def:L0}
\mathcal{L}_0 W_0 := \Pdiff_0(\I \K) W_0 + W_0 \Pdiff_0^*(\I \K)=0.
\end{equation}
In this equation the dispersion matrix
\[
\Pdiff_0(\I \K) = - \begin{pmatrix}
0 & c_0^2 \\
\F(\I \K) & 0
\end{pmatrix},
\]
admits the following spectral representation
\begin{equation*}
\Pdiff_0 = \lambda_+\B_+\C_+^\ast + \lambda_-\B_-\C_-^\ast, 
\end{equation*}
where
\begin{equation}
\label{def:c+}
\lambda_\pm(\K) := \pm  \I c_0 q_0(\I\K), \qquad \B_\pm(\K):= \frac{1}{\sqrt{2}}\begin{pmatrix} \pm \I q_0^{-1}(\I \K) \\ c_0^{-1} \end{pmatrix} \quad \mathrm{and} \quad \C_\pm(\K):= \frac{1}{\sqrt{2}}\begin{pmatrix} \pm \I q_0(\I \K) \\ c_0 \end{pmatrix},
\end{equation} 
with 
\[\B_\pm^\ast \C_\pm=1.\]
Note that here, the $\pm$ signs are not related to the up- and down going waves introduced in Sect. \ref{sec:intro}. Using that $\left(\B_+(\K), \B_-(\K)\right)$ forms a basis of $\mathbb{R}^2$, the matrix $W_0$ itself can be decomposed as  
\begin{equation}
\label{def:a+}
W_0 = \sum_{j,l=\pm}\ed{jl} \B_j\B_l^* \qquad\text{with}\qquad \ed{jl} := \C_j^\ast W_0 \C_l. 
\end{equation}
Plugging this relation into \eqref{def:L0} gives $\ed{jl}=0$ for $j\neq l$, and then
\begin{equation}
\label{eq:dec_W0}
W_0 =\ed + \B_+\B_+^* +\ed - \B_-\B_-^*,\qquad\text{with}\qquad \ed +:= \ed{++}\qquad\text{and}\qquad \ed -:= \ed{--}.
\end{equation}
Also, using that $\C (\K)= \C (-\K)$ for any $\K$, we have
\begin{equation}\label{eq:rel_sym_a}
\ed \pm(\K)= \ed \mp(-\K),
\end{equation}
so that we just have to focus our attention on $\ed +$.

% ----------------------------------------------------------------------

\paragraph{First order term $W_1$}

Equating like powers of $\epsilon^{1/2}$ in \eqref{Wigner equation rescale}, together with \eqref{asymptotic expansion 1} and \eqref{asymptotic expansion0}, we obtain the following relation
\begin{equation*}
\Pdiff_0\left(\I\K + \frac{\mathbf{D}_\y}{2}  \right )W_1 + W_1 \Pdiff_0^*\left(\I\K - \frac{\mathbf{D}_\y}{2}  \right ) +  \mathcal{K} K W_0 +  \mathcal{K}^{*} W_0 K^* = 0.
\end{equation*}
To avoid singular terms and to preserve causality, a regularization term $\theta$ is added following \cite{Bal2005, Ryzhik1996} and will be sent to $0$ later on: 
\begin{equation}
\label{First order corrector}
\Pdiff_0\left(\I\K + \frac{\mathbf{D}_\y}{2}  \right )W_1 + W_1 \Pdiff_0^*\left(\I\K - \frac{\mathbf{D}_\y}{2}  \right )  + \theta W_1+  \mathcal{K} K W_0 +  \mathcal{K}^{*} W_0 K^* = 0.
\end{equation}
Now, taking the Fourier transform of \eqref{First order corrector} in $\y$ leads to 
\begin{equation}
\label{Fourier first order corrector}
\Pdiff_0\left(\I\K + \I\frac{\p}{2}  \right )\widehat{W_1} +  \widehat{W_1}\Pdiff_0^*\left(\I \K - \I \frac{\p}{2}  \right ) + \theta \widehat{W_1}  \\
+  \widehat{\V} (\x,\p) K W_0\left( \K + \frac{\p}{2}   \right) + \widehat{\V}(\x,\p) W_0\left( \K - \frac{\p}{2}   \right) K^* = 0,
\end{equation}
and using that $\left(\B_+(\K),  \B_-(\K) \right)$ forms a basis of $\mathbb{R}^2$ for all $\K$, $\widehat{W_1}$ can be decomposed as 
\begin{equation}\label{def:hatW1}
\widehat{W_1}(\p,\K)=\sum_{j,l=\pm}\alpha_{jl}(\x,\p,\K)\B_j\left(\K+ \frac{\p}{2}\right) \B_l^*\left(\K- \frac{\p}{2}\right).
\end{equation}
Projecting \eqref{Fourier first order corrector} on the left on $\C_j^*\left(\K+\frac{\p}{2}\right)$ and on the right on $\C_l\left(\K-\frac{\p}{2}\right)$, we obtain
\begin{equation}
\label{def:alpha}
\alpha_{jl}(\x,\p,\K)=\frac{\widehat{\V}(\x,\p)}{2c_0^2} \left(\frac{\lambda_j\left( \K + \p/2   \right)\ed l\left( \K - \p/2   \right)   -         \lambda_l\left( \K - \p/2   \right)  \ed j\left( \K + \p/2   \right)     }{\lambda_j\left( \K + \p/2   \right) - \lambda_l\left( \K - \p/2   \right)    +\theta  }\right),
\end{equation} 
where we have used the three following relations 
\begin{equation*}
%\label{eq:formulas}
\lambda_\pm^* = - \lambda_\pm, \qquad \B_l^*(\p)K^*\C_j(\q)=\frac{1}{2c_0^2}\lambda_j(\q),\qquad \text{and}\qquad \C_j^*(\p)K\B_l(\q)=-\frac{1}{2c_0^2}\lambda_j(\p).
\end{equation*}

%—————————————————————————————————————————
%Radiative transfer equation
%—————————————————————————————————————————
         
\subsubsection{Radiative transfer equation}
\label{Radiative transfer equation}

To conclude and derive the radiative transfer equation we have to discuss the second order term $W_2$ that appears by equating powers of $\epsilon$ in \eqref{Wigner equation rescale}, together with \eqref{asymptotic expansion 1} and \eqref{asymptotic expansion0}. This way, we obtain
\begin{equation}
\label{Second order}
\Pdiff_0\left( \I \K + \frac{\mathbf{D}_\y}{2}   \right)W_2 +W_2 \Pdiff_0^*\left( \I \K - \frac{\mathbf{D}_\y}{2}   \right) + \mathcal{K} K W_1+  \mathcal{K}^{*} W_1 K^* \\
+ \partial_t W_0 + \Pdiff_1(\I\K)W_0 + W_0 \Pdiff_1^*(\I \K) =0,
\end{equation}
where the last terms do not depend on $\mathbf{D}_\y$ because $W_0$ does not depend on the $\y$-variable. Thanks to the decomposition \eqref{eq:dec_W0} the term $W_2$ can be chosen as being orthogonal to $W_0$ ($W_0$ is expanded over a two dimensional basis in a four dimensional vector space for any fixed $\K$) so that we necessarily have 
\[
\C^*_+(\K)W_2(t,\x,\y,\K)\C_+(\K) = 0.
\]
As a result, projecting \eqref{Second order} on the left on $\C^*_+(\K)$ and on the right on $\C_+(\K)$, we obtain 
\begin{equation}
\label{eq:ed0}
\partial_t \ed + + \mathcal{L}_1 W_1 + \mathcal{L}_2 W_0 = 0,
\end{equation}
with
\[
\mathcal{L}_1W_1(\K) :=  \C^*_+(\K)( \mathcal{K} K W_1 + \mathcal{K}^{*} W_1 K^*) \C_+(\K),
\]
and 
\[
\mathcal{L}_2W_0(\K) :=  \C^*_+(\K)\big(\Pdiff_1(\I\K)W_0 + W_0 \Pdiff_1^*(\I \K)\big)\C_+(\K) = c_0 \nabla_\K q(i\K)\cdot \nabla_\x \ed + = c_0 \widehat{\K} \cdot \nabla_\x \ed +,
\]
remembering that $\Pdiff_1$ is defined through (\ref{asymptotic expansion0} - \ref{asymptotic expansion}).

Regarding the term $\mathcal{L}_1W_1$, considering \eqref{def:hatW1} and~\eqref{def:alpha}, we can factorize $\widehat{W_1}$ as
\begin{equation}\label{def:calW1}
\widehat{W_1}(\x,\p,\K)= \widehat{\V}(\x,\p) \W_1(\p,\K),
\end{equation}
and obtain
\begin{align*}
\widehat{\mathcal{L}_1W_1}(\x,\p, \K) & = \int \widehat{\V}(\x,\R)\widehat{\V}(\x,\p-\R)  \\
 & \times \C^*_+(\K)  \Big(K\W_1 \Big(\p-\R,\K-\frac{\R}{2}\Big)  + \W_1\Big(\p-\R,\K+\frac{\R}{2}\Big)K^* \Big)\C_+(\K)\frac{d\R}{(2\pi)^3}.     
\end{align*}
At this step we invoke a mixing argument on $\widehat{\V}$ defined through a mixing property on $\widehat V$. This step of averaging can be justified rigorously following the approach of \cite{butz2015}. Following this work, we could even prove a self-averaging property (that is the following result would hold for $\ed +$ and not only for $\langle \ed +\rangle $ as it is described below). Nevertheless, this approach being very technical, with more involved mathematics, is beyond the scope of this paper, and we choose a more formal derivation. Assuming the following mixing property for $\widehat V$
\begin{equation}\label{eq:mixing}
\big\langle\widehat{V}(\mathbf{s}_1)\widehat{V}(\p-\mathbf{s}_2) U(\p, \mathbf{s}_1', \mathbf{s}'_2)\big\rangle = \big\langle\widehat{V}(\mathbf{s}_1)\widehat{V}(\p-\mathbf{s}_2) \big\rangle \big\langle U(\p, \mathbf{s}_1', \mathbf{s}'_2)\big\rangle, 
\end{equation}
and using that
\[
\widehat \V (\x,\p) = \widehat V(\p)\mathbf{1}_{\mathbb{R}_-}(\xn) + \widehat V(\pperp,-p_n)\mathbf{1}_{\mathbb{R}_+}(\xn)
\]
is a linear combination of two $\widehat V$, we then write
\begin{align*}
\big\langle \widehat{\mathcal{L}_1W_1}&(\x,\p, \K) \big\rangle  = \int \Big\langle \widehat{\V}(\x,\R)\widehat{\V}(\x,\p-\R) \Big\rangle \\
 & \times \C^*_+(\K)  \Big(K\Big\langle\W_1 \Big(\p-\R,\K-\frac{\R}{2}\Big)\Big\rangle  + \Big\langle \W_1\Big(\p-\R,\K+\frac{\R}{2}\Big) \Big\rangle K^* \Big)\C_+(\K)\frac{d\R}{(2\pi)^3}.   
\end{align*}
Here we apply the mixing relation w.r.t. the variable $\p$ of $\widehat \V $, which is the Fourier variable of the fast variable $\y$ to justify formally this relation. In this later relation (as well as in \eqref{eq:mixing}) we make an abuse of notation by using an equal sign, while this relation would hold in the limit $\epsilon\to0$. 

As a result, averaging \eqref{eq:ed0} yields
\begin{equation}\label{eq:RTE_tmp}
\partial_t \langle \ed + \rangle + c_0 \widehat{\K} \cdot \nabla_\x \langle \ed + \rangle + \big\langle\mathcal{L}_1W_1\big\rangle = 0,
\end{equation}
with
\begin{equation}
%\label{eq:dec_L1}
\big\langle \widehat{\mathcal{L}_1W_1}(\x,\p, \K)  \big\rangle = c_0^4 \delta(\p) \left(   \int\widehat{R}(\R) K \left< \W_1\left(\p-\R,\K-\frac{\R}{2}\right) \right> d\R \right. \nonumber\\
+ \left. \int \widehat{R}(\R) \left< \W_1\left(\p-\R,\K+\frac{\R}{2}\right) \right> K^* d\R \right),
\end{equation}
according to \eqref{Power spectra}. Considering now the change of variable $\R \to \K-\q$ for the term involving $K\W_1$ and $\R \to \q-\K$ for the one involving $\W_1 K^\ast$ yields 
\begin{equation}
%\label{Lambda}
\big\langle \widehat{\mathcal{L}_1W_1}(\x, \p, \K)  \big\rangle =   c_0^4\delta(\p) \int_{\mathbb{R}^3}\widehat{R}(\K-\q)\left[K \left<\W_1\left(\q-\K,\frac{\K+\q}{2}\right)\right> \right.\nonumber\\
 +\left.  \left<\W_1\left(\K-\q,\frac{\K+\q}{2}\right) \right>  K^*\right] d\q, 
\end{equation}
where we have used the relation $\widehat{R}(-\mathbf{r})=\widehat{R}(\mathbf{r})$. Going back to the definition of $\W_1$, given by \eqref{def:calW1} through \eqref{def:hatW1} and~\eqref{def:alpha}, and sending the regularization term $\theta$ to $0$, knowing that in the sense of distributions 
\begin{equation*}
\frac{1}{\I x + \theta} \, \underset{\theta \searrow 0}{\longrightarrow} \, \frac{1}{\I x} + \pi\delta(x),
\end{equation*}
we obtain
\begin{align*}
\int_{\mathbb{R}^3} \e^{\I \y \cdot \p} \big\langle \widehat{\mathcal{L}_1W_1}(\x, \p, \K)  \big\rangle \frac{d\p}{(2\pi)^3} \underset{\theta \searrow 0}{\longrightarrow} \Sigma(\K) & \langle \ed + \rangle(\K) \\
&- \int_{\mathbb{R}^3}\sigma(\K,\q)\langle \ed + \rangle(\q)\delta(c_0(\vert \q \vert -\vert\K\vert ))d\q,
\end{align*}
with
\begin{equation*}
%\label{scattering coefficiant 1}
\Sigma(\K)=\frac{\pi c_0^2\vert \K \vert^2}{2(2\pi)^3}\int_{\mathbb{R}^3} \widehat{R}(\K-\q) \delta(c_0(\vert \q \vert -\vert\K\vert ))d\q, 
\end{equation*}
and 
\begin{equation*}
%\label{scattering coefficiant 2}
\sigma(\K,\q)=\frac{\pi c_0^2\vert \K \vert^2}{2(2\pi)^3}\widehat{R}(\K-\q).
\end{equation*}
Finally, passing to the limit $\theta\searrow 0$ in \eqref{eq:RTE_tmp} yields
\begin{multline}
\label{equation de transport}
\partial_t \langle \ed + \rangle(t,\x,\K)  + c_0 \widehat{\K} \cdot \nabla_\x\langle \ed + \rangle(t,\x,\K)   = -\Sigma(\K)\langle \ed + \rangle(t,\x,\K)  \\
+\int_{\mathbb{R}^3}\sigma(\K,\q)\langle \ed + \rangle(t,\x,\q) \delta(c_0(\vert \q \vert -\vert\K\vert ))d\q,
\end{multline}
equipped with the initial condition
\begin{equation}\label{eq:a0}
\ed + (t=0,\x,\K) = \frac{1}{2}\Big\vert c_0^{-1} \widehat{B}(\K)  -i \vert \K \vert \widehat{A}(\K) \Big\vert^2\delta(\x-\x_0),
\end{equation}
according to \eqref{eq:cond_init}, \eqref{def:c+}, and \eqref{def:a+}. As a result, despite the lack of stationarity for $\V$ w.r.t. the $x_n$-variable the effective scattering phenomena for the waves energy propagation can be described by the same RTE as for fully stationary random media \cite{Bal2005}. Also, as already mentioned, following the strategy of \cite{butz2015}, we could have proved that $\ed +$ itself is the solution to the deterministic equation \eqref{equation de transport}, leading to the relation
\[
\ed + = \langle \ed + \rangle.
\]
Keeping in mind the relation \eqref{def:a+} and \eqref{eq:dec_W0}, we drop the $+$ dependence for $\ed +$ in the remaining of the paper, and only use the notation $a$ to avoid any confusion with the up- and down-going waves introduced in Sect. \ref{sec:intro}.

To summarize, remembering \eqref{asymptotic expansion 1}, \eqref{eq:dec_W0}, and \eqref{eq:rel_sym_a}, the asymptotic behavior for the Wigner transform $W^{--}_\epsilon$ is then given by
\begin{equation}
\label{def:W0--}
W^{--}_0(t,\x,\K) := \lim_{\epsilon\to 0}W^{--}_\epsilon(t,\x,\K) = a(t,\x, \K) \BB(\K)  + a(t,\x,-\K)\BB^T(\K),
\end{equation}
where
\begin{equation}\label{def:B}
\BB(\K):= \frac{1}{2}\begin{pmatrix}
1/\vert \K\vert^2 & i/(c_0\vert \K\vert) \\
- i/(c_0\vert \K\vert) & 1/c_0^2,
\end{pmatrix},
\end{equation}
and $\BB^T$ stands for its transposition.

%—————————————————————————————————————————
%Energy contribution of the cross terms
%—————————————————————————————————————————
 
\subsection{Energy contribution of the cross terms}
\label{Energy contribution of the cross terms}

From \eqref{Wigner total} and \eqref{sym wigner_cross}, to complete the analysis of $W_\epsilon$, it remains to describe the asymptotic behavior of $W^{-+}_\epsilon$. This term describes the correlations between the up- and down- going waves produced on each side of the interface $\{x_n=0\}$. 

%—————————————————————————————————————————
%RTE for the cross-Wigner transform
%—————————————————————————————————————————

\subsubsection{RTE for the cross-Wigner transform}

Following the very same steps of Sect.~\ref{RTE for the self-Wigner transforms}, it can be shown that $W^{-+}_\epsilon$ can be described asymptotically by a standard (linear) RTE. However, in the case of initial conditions far away from the border, the initial condition for $W^{-+}_\epsilon$ goes to $0$ as $\epsilon\to 0$. Indeed,
\begin{align}\label{eq:cross_init}
W^{-+}_{\epsilon}(t=0,\x,\K)&=\int_{\mathbb{R}^3}\e^{\I\K\cdot \y}  \U^-_\epsilon\left(t=0,\x-\epsilon\y/2\right)  \U^{+*}_\epsilon\left(t=0,\x+ \epsilon\y/2\right) \frac{d\y}{(2\pi) ^3}\\
& =\frac{1}{\epsilon^3} \int_{\mathbb{R}^3}\e^{\I\K\cdot \y}   \mathbf{w}_{1,\epsilon} \left(\frac{\x}{\epsilon}- \frac{\y}{2} -\frac{\x_0}{\epsilon}\right)\nonumber\\
&\hspace{1cm}\times  \mathbf{w}_{2,\epsilon}^*\left(\frac{\xperp}{\epsilon} + \frac{\yperp}{2} -\frac{\x_{0,\perp}}{\epsilon},  \frac{x_n}{\epsilon}+ \frac{y_n}{2} +\frac{x_{0,n}}{\epsilon}        \right) \frac{d\y}{(2\pi) ^3}\nonumber\\
&= \int_{\mathbb{R}^3}\e^{ \I \x \cdot \mathbf{s}  }  \e^{-\I \x_{0,\perp} \cdot \mathbf{s}_\perp  } \e^{-2 \I x_{0,n} k_n /\epsilon  }  \widehat{\mathbf{w}}_{1,\epsilon} \left(\K + \frac{\epsilon \mathbf{s}}{2}\right)  \widehat{\mathbf{w}}_{2,\epsilon}^*\left(\K - \frac{\epsilon \mathbf{s}}{2}\right) d\mathbf{s},\nonumber
\end{align}
where 
\begin{equation}\label{def:w}
 \mathbf{w}_{1,\epsilon} (\x) :=\begin{pmatrix}  A(  \x  ) \\  c_\epsilon^{-2}(\x+\x_{0} ) B( \x ) \end{pmatrix} \quad \mathrm{and} \quad  \mathbf{w}_{2,\epsilon} (\x):= \begin{pmatrix}  A(  \x ) \\  c_\epsilon^{-2}( \xperp+\x_{0,\perp}, x_n-x_{0,n}) B(  \x ) \end{pmatrix},
\end{equation}
so that
\[
\lim_{\epsilon\to 0}W^{-+}_{\epsilon}(t=0,\x,\K)=0,
\]
in a weak sens, because of the fast phase $\e^{-2 \I k_n x_{0,n} /\epsilon}$ together with the Riemann-Lebesgue theorem. Therefore, the RTE would have null initial conditions leading to 
\[
\lim_{\epsilon\to 0} W^{-+}_{\epsilon}(t,\x,\K)=0,
\]  
for any time $t>0$. To describe the contribution provided by $W^{-+}_{\epsilon}$ we need a more careful approach. 

%—————————————————————————————————————————
%Amplification at the interface for the cross-Wigner transform
%—————————————————————————————————————————

\subsubsection{Amplification at the interface for the cross-Wigner transform}

Remembering~\eqref{Wigner total} and~\eqref{sym wigner_cross}, the asymptotic total energy density is given by
\begin{align}\label{eq:Wtot}
W^{tot} (t,\x,\K) &:= \lim_{\epsilon\to 0} W_\epsilon (t,\x,\K) = W^{--}_0(t,\x,\K) + W^{--}_0(t,(\xperp,-x_n),(\Kperp,-k_n))\\
& = (a(t,\x,\K) + a(t,(\xperp,-x_n),(\Kperp,-k_n)))\BB(\K)\nonumber\\
& + (a(t,\x,-\K) + a(t,(\xperp,-x_n),(-\Kperp,k_n)))\BB^T(\K), \nonumber
\end{align}
yielding the boundary condition \eqref{reflecting boundary conditions} at $\{x_n=0\}$ corresponding to the one derived in \cite{Ryzhik1997}, and the total energy contributions from all directions
\begin{align}\label{eq:energy1}
E(t,\x) &:= \int_{\mathbb{R}^3} W^{tot} (t,\x,\K) d\K = \int_{\mathbb{R}^3} W^{--}_0(t,\x,\K) + W^{--}_0(t,(\xperp,-x_n),\K) d\K,\\
& =\int_{\mathbb{R}^3} (a(t,\x,\K) + a(t,(\xperp,-x_n),\K)) \DD(\K) d\K, \nonumber
\end{align}
where
\begin{equation*}%\label{def:D}
\DD(\K):= \begin{pmatrix}
1/\vert \K\vert^2 & 0 \\
0 & 1/c_0^2,
\end{pmatrix}.
\end{equation*}

To exhibit the nontrivial contribution to $W^{-+}_{\epsilon}$, we introduce the following \emph{shifted} wave equation 
\begin{equation*}
%\label{equation onde translate}
\partial^2_{tt} q^\epsilon(t,\x)-\ctilde^2\left(\frac{\x_\perp}{\epsilon},\frac{x_n-x_{0,n}}{\epsilon}\right) \Delta q^\epsilon(t,\x)  = 0 \qquad (t,\x)\in\mathbb{R}_+^\ast \times \mathbb{R}^3,\\
\end{equation*} 
equipped with the initial conditions 
\begin{equation*}
q^\epsilon(t=0,\x)=\frac{1}{\epsilon^{3/2}}A\left(  \frac{\x_\perp-\x_{0,\perp}}{\epsilon}, \frac{x_n}{\epsilon}   \right),
\end{equation*}
and
\begin{equation*}
\epsilon \partial_t q^\epsilon(t=0,\x)=\frac{1}{\epsilon^{3/2}} B\left(  \frac{\x_\perp-\x_{0,\perp}}{\epsilon}, \frac{x_n}{\epsilon}   \right).
\end{equation*}  
Considering now the vector field 
\begin{equation*}
\mathbf{v}_\epsilon(t,\x):=\begin{pmatrix}  q^\epsilon(t,\x) \\ \epsilon \, \ctilde^{-2}\left(\frac{\x_\perp}{\epsilon},\frac{x_n-x_{0,n}}{\epsilon}\right) \partial_t q^\epsilon(t,\x)  \end{pmatrix},
\end{equation*} 
it turns out that
\begin{equation*}
\U^-_\epsilon(t,\x_\perp,x_n)=\mathbf{v}_\epsilon(t,\x_\perp,-x_n+x_{0,n})     \quad \mathrm{and} \quad \U^+_\epsilon(t,\x_\perp,x_n)=\mathbf{v}_\epsilon(t,\x_\perp,x_n+x_{0,n}).
\end{equation*}
In this way, $W_\epsilon^{-+}$ can be rewritten as
\begin{align*}
W^{-+}_{\epsilon}(t,\x,\K)&=\int_{\mathbb{R}^3}\e^{\I\K\cdot \y}  \U^-_\epsilon\left(\x-\epsilon\y/2\right)  \U^{+*}_\epsilon\left(\x+ \epsilon\y/2\right) \frac{d\y}{(2\pi) ^3}\\
& = \int_{\mathbb{R}^3}\e^{\I\K\cdot \y}  \mathbf{v}_\epsilon\left(\x_\perp-\epsilon\y_\perp/2,x_{0,n}   + \epsilon y_n/2  - x_n   \right) \\
& \hspace{3cm} \times\mathbf{v}_\epsilon^*\left(\x_\perp + \epsilon\y_\perp/2,  x_{0,n}   + \epsilon y_n/2  + x_n  \right) \frac{d\y}{(2\pi) ^3}\\
& =  \frac{1}{2\pi} \int_{\mathbb{R}} \e^{\I k_ny_n}  \left( \int_{\mathbb{R}} \e^{-2\I p_n x_n/\epsilon} \W_\epsilon\left(t,  \x_\perp, x_{0,n} + \epsilon y_n/2      , \Kperp,p_n \right) dp_n \right)dy_n,       
\end{align*} 
where 
\begin{equation*}
\W_\epsilon(t,\x,\K):=\frac{1}{(2\pi)^3} \int_{\mathbb{R}^3} \e^{\I\K\cdot\y} \mathbf{v}_\epsilon(t,\x-\epsilon \y/2)  \mathbf{v} _\epsilon^*(t,\x+\epsilon\y/2).
\end{equation*}
We thus obtain the following expression for $W^{-+}_{\epsilon}$,
\begin{equation*}
W^{-+}_{\epsilon}(t,\x,\K) =  \frac{2}{\epsilon(2\pi)} \int_{\mathbb{R}} \int_{\mathbb{R}} \e^{2\I k_n(y_n-x_{0,n})/\epsilon}  \e^{-2\I p_n x_n/\epsilon} \W_\epsilon(t,  \xperp, y_n    , \Kperp,p_n ) dp_ndy_n.       
\end{equation*} 
The presence of highly oscillatory terms, $\e^{2\I k_n(y_n-x_{0,n})/\epsilon}$ and $\e^{-2\I p_n x_n/\epsilon}$, suggests to place our attention at the vicinity of the interface $\{ \xn = 0\}$ (within one wavelength), leading to the two changes of variables 
\[
\x \to \x_\epsilon=(\xperp,\epsilon \tilde{x}_n/2) \qquad \text{and} \qquad \K \to  \K_\epsilon =(\Kperp,\epsilon \tilde{ k}_n/2).
\]
The first change of variable $\x \to \x_\epsilon$ allows us to focus on phenomena near the boundary $\{ \xn = 0\}$, while the second one $\K \to \K_\epsilon$ allows to follow what propagates along this boundary. Finally, remembering that $\ctilde_\epsilon$ is even w.r.t. the $x_n$-variable, and that, thanks to~\eqref{eq:IC_p-+} and~\eqref{eq:IC_dp-+}, the initial conditions for the two fields are symmetrical with respect to $\xn=0$, we remark that
\[
\W_\epsilon(t,  \xperp, y_n    , \Kperp,p_n) = W_\epsilon^{- -}(t,  \xperp, x_{0,n} - y_n    ,\Kperp, p_n),
\]
so that
\[
W^{-+}_{\epsilon}(t,\x_\epsilon,\K_\epsilon) =  \frac{2}{\epsilon(2\pi)} \int_{\mathbb{R}} \int_{\mathbb{R}} \e^{-\I \tilde{k}_n y_n}  \e^{-\I p_n \tilde{x}_n} W_\epsilon^{- -}(t,  \xperp, y_n,\Kperp, p_n) dp_ndy_n. 
\]
As a result, the asymptotic (nontrivial) contribution of $W^{-+}_{\epsilon}$ can be described as
\begin{equation}
\label{asymptotic limit of cross terms}
\lim \limits_{\epsilon\to 0}\epsilon  W^{-+}_{\epsilon}(t, \x_\epsilon, \K_\epsilon)  = \frac{2}{(2\pi)} \int_{\mathbb{R}} \int_{\mathbb{R}} \e^{-\I \tilde{k}_ny_n}  \e^{-\I p_n \tilde{x}_n} W_0^{- -}\left(t,  \xperp, y_n    , \Kperp,p_n \right) dp_ndy_n, 
\end{equation}
where $W_0^{- -}$ is given by \eqref{def:W0--}. One can remark here that the boundary effects will not provide any additional numerical effort to be simulated. They can easily be obtained from $W_0^{- -}$ through the numerical simulation of $a$.  

Regarding now the energy contributions from all directions at the vicinity (within one wavelength) of the boundary $\{x_n=0\}$, we have 
\begin{align*}
E_{boundary}(t,\xperp,\tilde {x}_n) &:= \int_{\mathbb{R}^3} \lim_{\epsilon\to 0}W_\epsilon(t,\x_\epsilon,\K) d\K\\
& = \int_{\mathbb{R}^3} \lim_{\epsilon\to 0}W^{--}_\epsilon(t,\x_\epsilon,\K) + W^{--}_\epsilon(t,(\xperp,-\epsilon \tilde x_n/2),\K) d\K\\
& + \frac{1}{2} \int_{\mathbb{R}^3} \lim_{\epsilon\to 0} \epsilon W^{-+}_\epsilon(t,\x_\epsilon,\K_\epsilon) + \epsilon W^{-+}_\epsilon(t,(\xperp,-\epsilon \tilde x_n/2),\K_\epsilon) d\K,
\end{align*}
after the change of variable $\K\to \K_\epsilon$ for the integral of the cross terms. Finally, using \eqref{asymptotic limit of cross terms}, we obtain
\[
E_{boundary}(t,\xperp,\tilde {x}_n) = 2 \int_{\mathbb{R}^3} a(t, (\xperp,0),\K)(1+\cos(k_n \tilde{x}_n)) \DD(\K) d\K,
\]
exhibiting the effects of the boundary, and then
\[E_{boundary}(t,\xperp,\tilde {x}_n = 0) = 2 \, E(t,(\xperp,0)),\]
that is a doubling of the energy w.r.t. \eqref{eq:energy1}.

%—————————————————————————————————————————
%Time and space evolution of the boundary effects
%—————————————————————————————————————————

\subsection{Time and space evolution of the boundary effects}

In this section we investigate in more details the time and space evolution of the boundary effects \eqref{asymptotic limit of cross terms} along the interface $\{x_n=0\}$ by looking at
\[
\W_0(t, \xperp,\tilde x_n, \Kperp)  := \int_\mathbb{R} \lim_{\epsilon\to 0} W_\epsilon(t,\x_\epsilon,\Kperp,k_n)dk_n, 
\]
which can be rewritten thanks to \eqref{Wigner total}, \eqref{sym wigner_self}, \eqref{sym wigner_cross}, and \eqref{asymptotic limit of cross terms} as
\begin{align*}
\W_0&(t,  \xperp,\tilde x_n, \Kperp)  = \int_{\mathbb{R}} \lim_{\epsilon\to 0}W^{--}_\epsilon(t,\x_\epsilon,(\Kperp,k_n)) + W^{--}_\epsilon(t,(\xperp,-\epsilon \tilde x_n/2),(\Kperp,k_n)) dk_n \\
& + \frac{1}{2} \int_{\mathbb{R}} \lim_{\epsilon\to 0} \epsilon W^{-+}_\epsilon(t,\x_\epsilon,\Kperp,\epsilon \tilde k_n/2) + \epsilon W^{-+}_\epsilon(t,(\xperp,-\epsilon \tilde x_n/2),(\Kperp, -\epsilon \tilde k_n/2)) dk_n \\
& = 2\int_{\mathbb{R}} (1+\cos(k_n\tilde x_n)) W^{--}_0(t,(\xperp, 0),(\Kperp,k_n))dk_n\\
& = 2\int_{\mathbb{R}} (1+\cos(k_n\tilde x_n)) (a(t,(\xperp,0),(\Kperp,k_n)) \BB(\Kperp,k_n) \\
& \hspace{4cm}+a(t,(\xperp,0),(-\Kperp,k_n)) \BB^T(\Kperp,k_n)) dk_n.
\end{align*}
To describe more closely $a$, let us rewrite \eqref{equation de transport} under its mild formulation 
\begin{align*}
a(t,\x,\K) = & \,\e^{-\Sigma(\K)t} a_0(\x-c_0 t \widehat \K,\K) \\
& + \int_0^t \int \sigma(\K,\p) \e^{-\Sigma(\K)(t-s)} a(s,\x-c_0 \widehat\K(t-s),\p)d\p,
\end{align*}
where
\[
a_0(\x,\K):= a(t=0,\x,\K) = \mathbb{A}(\K)\delta(\x-\x_{0}),
\]
with $\mathbb{A}$ defined according to \eqref{eq:a0}. Iterating this relation yields the following Duhamel expansion 
\begin{align}\label{eq:iter_a}
a(t,\x,\K) &= \sum_{N\geq 0}\int_{\Delta_N(t)}d\mathbf{s}^{(N)} \int d\p^{(N)} \e^{-\sum_{j=0}^N\Sigma(\p_j)(s_j-s_{j+1})} \prod_{j=1}^N\sigma(\p_{j-1},\p_j) \\
&\times \mathbb{A}(\p_N) \delta\Big(\x-\x_0-c_0\sum_{j=0}^N \hat\p_j (s_j-s_{j+1})\Big)\nonumber,
\end{align}
where $\p^{(N)}:=(\p_1,\dots,\p_N)$, $\p_0:=\K$, and
\[
\Delta_N(t)=\{\mathbf{s}^{(N)}:=(s_1,\dots,s_N):\,0\leq s_N\leq \dots \leq s_1 \leq t\},
\]
with $s_0:=t$ and $s_{N+1}:=0$. In this expansion, the first term in the r.h.s ($N=0$) corresponds to the coherent component, while the other terms ($N\geq 1$) result from the multiple scattering. The $N$-th term in the sum corresponds to the component that undergoes $N$ scattering events. 

Now, inserting \eqref{eq:iter_a} into the definition of $\W_0$ yields
\[\W_0 = \W_{0}^c+\W_{0}^i,\]
where $\W_{0,c}$ contains the contribution of the coherent waves
\begin{align*}
\W_{0}^c(t,  \xperp,\tilde x_n, \Kperp) := 2\int  & dk_n \, (1+\cos(k_n \tilde x_n))  \e^{-\Sigma(\Kperp,k_n)t}\\
&\times \Big(\mathbb{A}(\Kperp,k_n)\delta(\x_\perp-\x_{0,\perp}-c_0 \widehat\p_{0,\perp}  t)\BB(\Kperp,k'_n(t))\\
&+\mathbb{A}(-\Kperp,k_n)\delta(\x_\perp-\x_{0,\perp}+c_0 \widehat\p_{0,\perp}  t)\BB^T(\Kperp,k'_n(t))\Big)\\
& \times \delta(x_{0,n} + c_0\widehat\p_{0,n} t),
\end{align*}
with the notation 
\[
\widehat \p_0 := \frac{(\Kperp,k_n)}{\sqrt{|\Kperp|^2 + k_n^2}},
\]
and $\W_{0}^i$ contains the contributions of the incoherent waves
\begin{align*}
\W_{0}^i(t, \xperp,& \tilde x_n, \Kperp) := 2\sum_{N\geq 1}\int_{\Delta_N(t)}d\mathbf{s}^{(N)} \int d\p^{(N)} \int dk_n \, (1+\cos(k_n \tilde x_n))\\
&\times \e^{-\sum_{j=0}^N\Sigma(\p_j)(s_j-s_{j+1})} \prod_{j=1}^N\sigma(\p_{j-1},\p_j) \\
&\times \Big(\mathbb{A}(\p_{N,\perp}, \p_{N,n}) \delta\Big(\x_\perp-\x_{0,\perp}-c_0\sum_{j=0}^N \widehat\p_{j,\perp}  (s_j-s_{j+1})\Big) \BB(\Kperp,k_n)\\
&+\mathbb{A}(-\p_{N,\perp}, \p_{N,n}) \delta\Big(\x_\perp-\x_{0,\perp}+c_0\sum_{j=0}^N \widehat\p_{j,\perp}  (s_j-s_{j+1})\Big)\BB^T(\Kperp,k_n) \Big) \\
&\times\delta\Big(x_{0,n} + c_0\sum_{j=0}^N \widehat\p_{j,n} (s_j-s_{j+1}) \Big).
\end{align*}
The last Dirac masses in $\W_{0}^c$ and $\W_{0}^i$ lead to the condition
\[
k_n=k'_n (t,\x_0) :=  |\Kperp|\frac{|x_{0,n}| }{\sqrt{c_0^2 t^2 - x_{0,n}^2}}\qquad \text{for}\qquad \W_{0,c},
\]
and
\[
k_n = \pm k'_{n}\Big(t-s_1,\tilde \x_0(\mathbf{s}^{(N)},\p^{(N)})\Big)  \qquad \text{for}\qquad \W_{0,i},
\]
where
\begin{equation}\label{def:tilde_x0}
\tilde  \x_{0}(\mathbf{s}^{(N)},\p^{(N)}) := \big(\tilde  \x_{0,\perp}(\mathbf{s}^{(N)},\p^{(N)}_\perp), \tilde  x_{0,n}(\mathbf{s}^{(N)},\p^{(N)}_n)\big) := \x_{0} + c_0\sum_{j=1}^N \widehat\p_{j} (s_j-s_{j+1}).
\end{equation}
Here, $\tilde  \x_{0}(\mathbf{s}^{(N)},\p^{(N)})$ reflects a position after $N$ scattering events. To compute this position, starting from $\x_0$, we follow successively the directions $\widehat\p_{j}$ for a time duration $s_j-s_{j+1}$. 

In $\W_{0}^c$ the later Dirac mass can then be rewritten as
\[
\delta(x_{0,n} + c_0 \hat\p_{0,n} t) = \delta(g(k_n))=\frac{1}{g'(k'_n)}\delta(k_n-k'_n),
\]
with 
\[
g'(k_n)=\frac{c_0 t |\Kperp|^2}{(|\Kperp|^2+k_n^2)^{3/2}},
\]
so that
\[
g'(k'_n)=\frac{c_0 t}{|\Kperp|}\Big(1-\Big(\frac{x_{0,n}}{c_0 t}\Big)^2\Big)^{3/2}.
\]
As a result, the contribution of the coherent waves can be explicitly given by
\begin{align*}
\W_{0}^c(t, \xperp,\tilde x_n, \Kperp)  = \big(1+\cos(k'_n(t,\x_0) \tilde x_n)\big)&\big(\mathbb{A}(\Kperp,k'_n(t,\x_0))\,\mathcal{I}_-(t,\x_0,\K_\perp)\\
& + \mathbb{A}(-\Kperp,k'_n(t,\x_0)))\,\mathcal{I}^T_+(t,\x_0,\K_\perp) \big),
\end{align*}
where
\[
\mathcal{I}_{\pm}(t,\x_0,\K_\perp):= \frac{2|\Kperp|\e^{-\Sigma(\Kperp,k'_n(t,\x_0))t}}{c_0 t\big(1-(\frac{x_{0,n}}{c_0 t}\big)^2\big)^{3/2}} \delta\Big(\x_\perp - \x_{0,\perp} \pm \sqrt{c_0^2 t^2-x_{0,n}^2}\widehat\K_\perp \Big)\BB(\Kperp,k'_n(t,\x_0)).
\]
The displacement of this contribution along the boundary with direction $\widehat\K_\perp$ is described through $\mathcal{I}_\pm$, each of them originate at $\x_{0,\perp}$ and propagate in opposite directions depending of the $\pm$ signs. The amplitudes of these contributions are affected by a geometric attenuation $1/(c_0t)$ and how the initial condition charges the current direction of the waves reaching the boundary $\mathbb{A}(\Kperp,k'_n(t,\x_0))$. These two contributions hold whether or not the propagation medium is heterogeneous. The difference comes from the attenuation term $\e^{-\Sigma(\Kperp,k_n(t,\x_0))t}$ depending on the current statistical properties of the medium heterogeneities. 

Regarding the contribution of the incoherent waves, the same strategy yields 
\begin{align}\label{eq:W_inc}
\W_{0}^i&(t, \xperp,\tilde x_n, \Kperp) = \sum_{N\geq 1}\int_{\Delta_N(t)}d\mathbf{s}^{(N)} \int d\p^{(N)} \\
& \e^{-\sum_{j=1}^N\Sigma(\p_j)(s_j-s_{j+1})}\prod_{j=1}^N\sigma(\p_{j-1},\p_j) \, \sigma\big((\Kperp,k'_n(t-s_1,\tilde \x_0(\mathbf{s}^{(N)},\p^{(N)}))),\p_1\big)\nonumber\\
&\times \big(1+\cos(k'_n(t-s_1,\tilde \x_0(\mathbf{s}^{(N)},\p^{(N)})) \tilde x_n)\big)\nonumber\\
&\hspace{3cm} \times \big(\mathbb{A}(\p_{N,\perp}, \p_{N,n}) \mathcal{I}_-(t-s_1,\tilde \x_0(\mathbf{s}^{(N)},\p^{(N)}),\K_\perp)\nonumber \\
&\hspace{3cm}+\mathbb{A}(-\p_{N,\perp}, \p_{N,n}) \mathcal{I}^T_+(t-s_1,\tilde \x_0(\mathbf{s}^{(N)},\p^{(N)}),\K_\perp)\big),\nonumber
\end{align}
after using the symmetries w.r.t. the $k_n$-variable and appropriate changes of variables.

One can observe here that the incoherent component provides a continuum of contribution w.r.t time through the integral over $\mathbf{s}^{(N)}$, and the three last lines of \eqref{eq:W_inc} provide a similar comportment as for the coherent component $\W_{0}^c$. However, the starting points of these contributions are described through the location $\tilde \x_{0}$, playing the role of fictitious sources w.r.t. the last scattering event. The position $\tilde \x_{0}$ is evaluated by \eqref{def:tilde_x0} and depends on all the previous scattering events (as for the overall contribution). Also, the time duration of each contribution holds on a time interval of length $t-s_1$ corresponding to the time of the last scattering event $s_1$ to the observation time $t$.   

%—————————————————————————————————————————
%Dirichlet boundary condition
%—————————————————————————————————————————

\subsection{Dirichlet boundary condition}
\label{Dirichlet boundary condition}

Under Dirichlet boundary conditions for the wave equation \eqref{equation des ondes}, the asymptotic analysis of the energy density is the same as for Neumann conditions, but with 
\begin{equation}\label{def:p_dirichlet}
\widetilde{p}^\epsilon(t,\x) = p_-^\epsilon(t,\x) - p_+^\epsilon(t,\x) \qquad (t,\x) \in  \mathbb{R}_+\times \mathbb{R}^3,
\end{equation}
instead of \eqref{superposition}, where $p_-^\epsilon$ and $p_+^\epsilon$ still satisfy \eqref{sym_p-p+}. This way, $\widetilde{p}^\epsilon$ is now odd with respect to the $x_n$-variable, which is compatible with Dirichlet boundary conditions. Despite the change of boundary conditions, the reflection conditions \eqref{reflecting boundary conditions} still hold true, as well as~\eqref{sym wigner_self} and~\eqref{sym wigner_cross}. Hence, similar results as \eqref{def:W0--} and \eqref{asymptotic limit of cross terms} can be obtained for the self- and cross-Wigner transforms respectively. In particular, \eqref{eq:energy1} still holds true. The difference lies in the decomposition \eqref{Wigner total}, which becomes here
\[
 W_\epsilon(t,\x,\K) = W_\epsilon^{--}(t,\x,\K) + W^{++}_\epsilon(t,\x,\K) - W^{-+}_\epsilon(t,\x,\K) - W_\epsilon^{+-}(t,\x,\K),
\]
because of the negative sign in \eqref{def:p_dirichlet}. Therefore, following the same lines as in Sect. \ref{Energy contribution of the cross terms} yields
\begin{equation*}%\label{asymptotic limit of cross terms Dirichlet}
E_{boundary}(t,\xperp,\tilde {x}_n) =  2 \int_{\mathbb{R}^3} a(t, (\xperp,0),\K)(1-\cos(k_n \tilde{x}_n)) \DD(\K) d\K.
\end{equation*}
One can see here a canceling of the energy intensities at the boundary,
\[E_{boundary}(t,\xperp,\tilde {x}_n = 0) = 0,\]
which is consistent with Dirichlet boundary conditions.

%—————————————————————————————————————————
%Asymptotic analysis with an initial condition close to the border
%—————————————————————————————————————————

\section{Asymptotic analysis with an initial condition close to the border}
\label{Asymptotic analysis with a source close to the border}

In this part, we assume an initial condition located close (within one wavelength) to the border. In other words, $\x_0$ becomes here 
\[\xoeps=(\x_{0,\perp} , \epsilon x_{0,n}).\]
In this situation, all the boundary effects described in Sect.~\ref{Energy contribution of the cross terms} still hold true, and the dramatic difference lies away from the boundary. In fact, going back to \eqref{eq:cross_init}
\begin{align*}
W^{-+}_{\epsilon}(t=0,\x,\K)&=\int_{\mathbb{R}^3}\e^{\I\K\cdot \y}  \U^-_\epsilon\left(t=0,\x-\epsilon\y/2\right)  \U^{+*}_\epsilon\left(t=0,\x+ \epsilon\y/2\right) \frac{d\y}{(2\pi) ^3}\\
& = \int_{\mathbb{R}^3}\e^{ \I \x \cdot \mathbf{s}  }  \e^{-\I \x_{0,\perp} \cdot \mathbf{s}_\perp  } \e^{-2 \I x_{0,n} k_n  }  \widehat{\mathbf{w}}_{1,\epsilon} \left(\K + \frac{\epsilon \mathbf{s}}{2}\right)  \widehat{\mathbf{w}}_{2,\epsilon}^*\left(\K - \frac{\epsilon \mathbf{s}}{2}\right) d\mathbf{s},
\end{align*}
where $\mathbf{w}_{1,\epsilon}$ and $\mathbf{w}_{2,\epsilon}$ are given by \eqref{def:w}, we obtain now a non-vanishing initial condition for the cross-Wigner transforms
\begin{equation}\label{eq:init_cond_cross}
\lim_{\epsilon\to 0}W^{-+}_{\epsilon}(t=0,\x,\K)=  \e^{-2 \I x_{0,n} k_n  } \widehat{S}(\K) \widehat{S}^\ast(\K)  \delta(\xperp - \x_{0,\perp}) \delta(x_n),
\end{equation}
taking place at the boundary $\{x_n=0\}$, and where $\widehat{S}$ is defined by \eqref{eq:cond_init}. As a result, following the very same lines as in Sect. \ref{Asymptotic analysis with a source far from the border}, one can write
\begin{align*}
W^{-+}_{0}(t,\x,\K) &:=\lim_{\epsilon\to 0}W^{-+}_{\epsilon}(t,\x,\K) \\
&=  a'(t,\x,\K)\BB(\K) + a'(t,\x,-\K)\BB^T(\K),
\end{align*}
where $a'$ satisfies the RTE \eqref{equation de transport} with initial condition given by \eqref{eq:init_cond_cross}, and $\BB$ defined by \eqref{def:B}. From \eqref{Wigner total}, \eqref{sym wigner_self} and~\eqref{sym wigner_cross}, the asymptotic of the full  Wigner transform is given by 
\begin{align*}
\widetilde W^{tot} (t,\x,\K) &:= \lim_{\epsilon\to 0} W_\epsilon (t,\x,\K) \\
& = W^{--}_0(t,\x,\K) + W^{--}_0(t,(\xperp,-x_n),(\Kperp,-k_n))\\
& \pm W^{-+}_0(t,\x,\K) \pm W^{-+}_0(t,(\xperp,-x_n),(\Kperp,-k_n))\\
& = (\tilde a(t,\x,\K) + \tilde a(t,(\xperp,-x_n),(\Kperp,-k_n)))\BB(\K)\\
& + (\tilde a(t,\x,-\K) + \tilde a(t,(\xperp,-x_n),(-\Kperp,k_n)))\BB^T(\K),
\end{align*}
where
\[\tilde a = a \pm a',\]
exhibiting the extra contribution $a'$ compared to $W^{tot}$ given by \eqref{eq:Wtot} for a source far from the border, and for which no boundary effect were taking place. Here, the $\pm$ signs depend on the boundary condition under consideration. A $+$ sign holds for Neumann boundary condition, and a $-$ sign for Dirichlet conditions. Note that $\tilde a$ satisfies the same RTE \eqref{equation de transport} as $a$, by linearity, but with initial conditions
\[
\tilde a(t=0,\x,\K) = 2\big(1 \pm \cos( 2 x_{0,n} k_n  )\big)a(t=0,\x,\K),
\]
reflecting the effects of the boundary. Sending $x_{0,n}$ to $0$ provides, for Neumann boundary condition, a doubling of the initial condition, so that 
\[\tilde a(t,\x,\K) = 2 \,a(t,\x,\K).\]
In this context, the boundary effects produce a doubling of the energy density that propagates all over the medium and not only located at the vicinity of the boundary. In other words, a source very close to the border with the Neumann boundary conditions produce some resonance effects allowing the initial condition to interact with itself and produce a doubling of the total propagating energy density,
\[
\widetilde W^{tot} = 2 \, W^{tot} 
\]
compared to a source located far away from the boundary \eqref{eq:Wtot}. Note that to respect the parity of the initial conditions $A$ and $B$ in \eqref{eq:cond_init_eps}, and the need to be compatible with the boundary conditions, the two functions $A$ and $B$ need to be sent to Dirac masses at $0$ when sending $x_{0,n}$ to $0$.     

For Dirichlet boundary conditions, sending $x_{0,n}$ to $0$ then yields
\[
\tilde a(t=0,\x,\K) = 0,
\]
and therefore a complete cancellation of the energy density
\[
\widetilde W^{tot} = 0, 
\]
which is consistent with this type of boundary conditions.

%—————————————————————————————————————————
%Conclusion
%—————————————————————————————————————————

\section{Conclusion}
\label{Conclusion}

In this paper, an asymptotic analysis has been provided allowing to derive a radiative transfer equation for acoustic waves energy propagating in a randomly fluctuating half-space, as well as describing explicitly the propagation of radiative interference located within one wavelength of the boundary. The approach has been built on the method of images, where the half-space problem has been extended to a full-space, and two symmetrical wave fields (the up and down-going waves) have been considered. The energy densities for these wave fields verify the same RTE \eqref{equation de transport}. Moreover, the interferences between the the up and down-going waves are responsible for contributions to the total energy, supported within one wavelength of the boundary. More specifically, at the border, under Neumann boundary conditions the interferences yields a doubling of the intensity and under Dirichlet boundary condition yields a canceling of the intensity. In the case of a source close (within one wavelength) to the border, the contribution due to the interference effects does not vanish far from the border, because the up and down-going waves evolve within one wavelength of each other and then produce interferences throughout the propagation domain.

%—————————————————————————————————————————
%
%—————————————————————————————————————————

\printbibliography

\end{document}